# PBR theorem and Einstein's quantum hole argument

Galina Weinstein

Upon reading Einstein's views on quantum incompleteness in publications or in his correspondence after 1935 (the EPR paradox), one gets a very intense feeling of deja-vu. Einstein presents a quantum hole argument, which somewhat reminds of the hole argument in his 1914 "Entwurf" theory of general relativity. PBR write in their paper, "An important step towards the derivation of our result is the idea that the quantum state is physical if distinct quantum states correspond to non-overlapping distributions for [physical states] λ". PBR then conclude, "The general notion that two distinct quantum states may describe the same state of reality, however, has a long history. For example, in a letter to Schrödinger containing a variant of the famous EPR (Einstein-Podolsky-Rosen) argument", and they refer to Einstein's quantum hole argument. This short paper discusses the PBR theorem, and the connection between the PBR argument and Einstein's argument and solution.

In philosophy we usually make a distinction between two views: the view that something corresponds directly to reality, *observation-independent*: refers to what is there, what is "physically real", and independent of anything we say, believe, or know about the system. In this case, quantum states represent knowledge about an underlying reality, real objects, real properties of quantum systems.

David Wallace explains that if we consider Schrödinger's cat *gedankenexperiment*, then from the above point of view, Schrödinger's cat in superposition of two states is a monster inside a box: both alive and dead cat at the same time. How come then when we open the box and observe (and measure) the both alive and dead cat at same time, we only see alive cat or dead cat? The orthodox interpretation of quantum mechanics invokes the collapse of the wave function whereby the act of observing the cat causes it to turn into alive-cat or dead-cat state. If the quantum state is a real physical state, then collapse is a mysterious physical process, whose precise time of occurrence is not well-defined. Accordingly, people who hold this view are generally led to alternative interpretations that eliminate the collapse, such as Everett's relative state formulation of quantum mechanics, many-worlds theory.[1]

The "state of knowledge" view, *observation-dependent*: refers to experimenter's knowledge or information about some aspect of reality; what we think, know, or believe what is reality. We presuppose tools of observation and measurement. In this case, the quantum state does not represent knowledge about some underlying reality, but rather it only represents experimenter's knowledge or information about some aspect of reality; his knowledge about the consequences of measurements that we might make on the system; it does not imply the existence of any real physical process.

From this anti-realist point of view, collapse of the wavefunction need be no more mysterious than just the instantaneous *Bayesian updating* of a probability distribution upon obtaining new information: Hence Schrödinger's cat is also not at all mysterious. When we consider superposition of states, both alive and dead cat at the same time, we mean that it has a fifty percent probability of being alive and a fifty percent probability of being dead (what is the likelihood of the cat being dead or alive). The process depends for its action on observations, measurements, and the knowledge of the observer. The collapse of the wave function corresponds to us observing and finding out whether the cat is dead or alive.[2]

Erhard Scheibe introduced the notions of *epistemic* and *ontic* states of a system. The *Ontic state* of the system is the system just the way it is, it is *empirically inaccessible*. It refers to individual systems without any respect to their observation or measurement. On the other hand, the *epistemic state* of the system *depends on observation and measurement*; it refers to the knowledge that can be obtained about an ontic state.[3]

As commonly understood, Bohr was advocating epistemic physics while Einstein was considering ontic physics. And indeed Jaynes wrote that the two physicists were not discussing the same physics: "needless to say, we consider all of Einstein's reasoning and conclusions correct on his level; but on the other hand we think that Bohr was equally correct on his level, in saying that the act of measurement might perturb the system being measured, placing a limitation on the information we can acquire and therefore on the predictions we are able to make. There is nothing that one could object to in this conjecture, although the burden of proof is on the person who makes it". Hence, the famous Bohr-Einstein debate was never actually resolved in favor of Bohr – although common thinking even among physicists and philosophers of science is that it was.[4]

Anton Zeilinger polled 33 participants of a conference on the foundations of quantum mechanics on their interpretations of quantum physics, and he found that philosophers of science indeed favored "Bohr". Zeilinger asked the participants: "Do you believe that physical objects have their properties well defined prior to and independent of measurement?" As to Einstein's view of quantum mechanics, he asked: "Is it correct?" No one thought it was. "Is it wrong?" 64% thought so. "Will it ultimately turn out to be correct?" 6% thought it would. Zeilinger's conclusion was: "In wording our question, we deliberately did not specify what exactly we took Einstein's view of quantum mechanics to be. It is well known, in fact, that Einstein held a variety of views over his lifetime. The overarching themes we were after, and the themes most people, we believe, would associate with Einstein – are a subtle flavor of realism, as well as the possibility of a deeper description of nature beneath quantum mechanics. Interestingly, none of the respondents brought himself to declaring Einstein's view as correct, although two people suggested that Einstein would ultimately be vindicated. One respondent sounded a conciliatory note. 'Einstein's view is wrong. But he still thought more clearly than anyone else in his time. There is still much to learn from him'." [5]

Zeilinger asked: "Bohr's view of quantum mechanics: Is it correct?" 21% thought it was correct. Zeilinger concluded: "Just as with the previous question about Einstein's view of quantum mechanics, we did not elaborate on the specifics of what we meant by 'Bohr's view.' Bohr, of course, has become associated with a variety of positions, and it is likely that in responding to the question, each participant had a slightly different set of ideas and slogans in mind." (Bohr's interpretations are different from what we call the Copenhagen interpretation).[6]

The million-dollar question is: What does a quantum state represent? What is the quantum state? Is the quantum wavefunction an ontic state or an epistemic state? Using the terminology of Harrigan and Spekkens, let us ask: is it possible to construct a *ψ-ontic* model? A *ψ-ontic hidden variable model* is a quantum state which is *ontic*, but we construct some underlying *ontic* hidden variable states theory. Hidden variable theories are always ontic states theories.

Or else, is it possible to construct a *ψ-epistemic* model? A *ψ-epistemic hidden variable model* is a quantum state which is *epistemic*, but there is some underlying *ontic* hidden state, so that quantum mechanics is the statistical theory of this ontic state.[7]

In a paper entitled, "On the Reality of the Quantum State" by Matt Pusey, Jon Barrett and Terry Rudolph (henceforth known as PBR), PBR answer the above question in the negative, ruling out ψ-epistemic theories, and attempting to provide a ψ-ontic view of the quantum state.[8]

PBR present a *no-go theorem* which is formulated for an *ontic* hidden variable theory: any model in which a quantum state represents mere information about an underlying physical state of the system must make predictions which contradict those of quantum theory. In the terminology of Harrigan and Spekkens, the PBR theorem says that *ψ-epistemic models cannot reproduce the predictions of quantum theory*.[9]

The PBR theorem holds only for systems that are prepared independently, have independent physical states (*independent preparations*). A system has an ontic hidden state λ. A quantum state ψ describes an experimenter's information which corresponds to a distribution of ontic hidden states λ. PBR show that when distinct quantum states ψ correspond to disjoint probability distributions of ontic hidden variables (i.e., what PBR call independent preparations, they do not have values of ontic hidden variables in common), these quantum states ψ are ontic and they are not mere information. PBR write that in this case the quantum state ψ can be inferred uniquely from the physical state of the system and hence satisfies the definition of a physical property. "Informally, every detail of the quantum state is 'written into' the real physical state of affairs". And if the states of a quantum system do not correspond to ontic disjoint probability distributions (the distributions of the values of ontic hidden values overlap), the quantum wavefunctions are said to be epistemic. "Our main result is" that for distinct quantum states ψ, if the distributions overlap, "then there is a contradiction with the predictions of quantum theory".[10]

The PBR theorem is in the same spirit as *Bell's no-go theorem*, which states that *no local* theory can reproduce the predictions of quantum theory. Bell's theorem was formulated for *ontic* hidden variable theory as well. Bell's theorem shows that a *ψ-epistemic* hidden variable theory which is *local* is forbidden in quantum mechanics, i.e, any ψ-epistemic hidden variable theory must be non-local in order to reproduce the quantum statistics of entanglement (EPR).

According to the EPR 1935 paper, it seems that Einstein favored a ψ-epistemic local interpretation. However, Einstein's correspondence after 1935 on EPR, or in his publications on EPR such as a 1936 paper "Physics and Reality", reveal that Einstein ruled out locality for *ψ-ontic* hidden variable theories. Indeed the only realistic interpretation of quantum states that could possibly be local is ψ-epistemic, but the ψ-ontic model must be non-local.[11]

The EPR paper was not written by Einstein. In a 1935 letter to Schrödinger Einstein wrote, "For reasons of language this [paper] was written by Podolsky after many discussions. But still it has not come out as well as I really wanted; on the contrary, the main point was, so to speak, buried by the erudition".[12]

Einstein wrote in "Physics and Reality":[13] Consider a mechanical system consisting of two partial systems A and B which interact with each other only during a limited time. ψ is the wavefunction before their interaction. One performs measurements on A and determines A's state. Then B's ψ function of the partial system B is determined from the measurement made, and from the ψ function of the total system. This determination gives a result which depends upon which of the observables of A have been measured (coordinates or momenta). That is, depending upon the choice of observables of A to be measured, according to quantum mechanics we have to assign different quantum states $ψ_B$ and $ψ_B'$ to B. These quantum states are different from one another.

In the above letter to Schrödinger Einstein explains,[14]

"After the collision, the real state of (AB) consists precisely of the real state A and the real state of B, which two states have nothing to do with one another. *The real state of B thus cannot depend upon the kind of measurement I carry out on A.* But then for the same state of B there are two (in general arbitrarily many) equally justified $\psi_B$, which contradicts the hypothesis of a one-to-one or complete description of the real state".

And he ends his 1936 paper by saying:[15] "Since there can be only one physical state of B after the interaction which cannot reasonably be considered to depend on the particular measurement we perform on the system A separated from B it may be concluded that the $\psi$ function is *not* unambiguously coordinated to the physical state. This coordination of several $\psi$ functions to the same physical state of system B shows again that the $\psi$ function cannot be interpreted as a (complete) description of a physical state of a single system. Here also the coordination of the $\psi$ function to an ensemble of systems eliminates every difficulty".

There is a resemblance between Einstein's Hole argument from the 1914 "Entwurf" theory of general relativity and Einstein's above views on incompleteness. In 1914 Einstein wrote:[16] "With respect to K there then existed different solutions G(x) and G'(x), which are different from one another, nevertheless at the boundary of the region both solutions coincide, *i.e., what is happening cannot be determined uniquely by generally-covariant differential equations for the gravitational field*." The correction should be that the reference system has no meaning and that the realization of two gravitational fields in the same region of the continuum is impossible.

Hence, we can call Einstein's above views on incompleteness by the name: *a quantum hole argument*; then according to Einstein, quantum physics is not ontic complete because for the same state of B there are two equally justified $\psi_B$.

In his *Autobiographical Notes* Einstein writes:[17] "If now the physicists A and B accept this reasoning as valid, then B will have to give up his position that the $\psi$-function constitutes a complete description of a real state. For in this case it would be impossible that two different types of $\psi$-functions could be assigned to the identical state of $S_2$".

PBR then suggest,[18]

"An important step towards the derivation of our result is the idea that the quantum state is physical if distinct quantum states correspond to non-overlapping distributions for $\lambda$ [the set of possible physical states that a system can be in…]".

Indeed the PBR argument:[19] "depends on few assumptions. One is that a system has a 'real physical state' – not necessarily completely described by quantum theory, but objective and independent of the observer. This assumption only needs to hold for systems that are isolated, and not entangled with other systems". It seems that the 1936 Einstein agrees with PBR.

Hence, "if the quantum state is a physical property of the system and apparatus, it is hard to avoid the conclusion that each macroscopically different component has a direct counterpart in reality".[20]

And PBR now present Einstein's quantum hole argument: [21]

"The general notion that two distinct quantum states may describe the same state of reality, however, has a long history. For example, in a letter to Schrödinger containing a variant of the famous EPR (Einstein-Podolsky-Rosen) argument, Einstein argues from locality to the conclusion that […]", and PBR cite the above paragraph from Einstein's 1935 letter to Schrödinger saying that for the same state of B there are two equally justified $\psi_B$".

They then further explain:[22]

"In this version of the argument, Einstein really is concerned with the possibility that there are two distinct quantum states for the same reality. He is not concluding that there are two different states of reality corresponding to the same quantum state (which would be the more commonly understood notion of incompleteness associated with Einstein)".

In a previous version of their ArXiv paper PBR speak more clearly: [23]

As far as we are aware, the precise formalisation of our question first appeared in Harrigan and Spekkens, where it is attributed to Hardy. But note that the general idea that two distinct quantum states may describe the same state of reality has a long history going back to Einstein. For example, in a letter to Schrödinger containing a variant of the famous EPR (Einstein-Podolsky-Rosen) argument, Einstein argues from locality to the conclusion that [… quoting Einstein's argument]. In this version of the argument, Einstein really is concerned with the idea that there are two distinct quantum states for the same reality, and not with the idea that there are two different states of reality corresponding to the same quantum state (the more commonly understood notion of incompleteness)".

However, the PBR theorem does not rule "Bohr" who believes in quantum physics that is *epistemic* complete (Wavefunctions are epistemic, but there is no underlying ontic hidden variable states theory[24]). Indeed PBR are aware of this possibility: "The PBR theorem only holds for systems that have 'real physical state' – not necessarily completely described by quantum theory, but objective and independent of the observer. This assumption only needs to hold for systems that are isolated, and not entangled with other systems. Nonetheless, this assumption, or some part of it, would be denied by instrumentalist approaches to quantum theory, wherein the quantum state is merely a calculational tool for making predictions concerning macroscopic measurement outcomes".[25]

It seems that the PBR theorem does not end the century-old debate about the ontology of quantum states. It does not prove, with mathematical certitude, that the ontic interpretation is right and the epistemic one is wrong.